\documentclass[letterpaper,10pt,conference]{ieeeconf}
\IEEEoverridecommandlockouts
\usepackage{cite}
\let\proof\relax

\usepackage{amsmath,amssymb,amsthm,mathtools,bm}
\usepackage{graphicx}
\usepackage{xcolor}
\graphicspath{{figures/}}
\let\labelindent\relax
\usepackage{booktabs,multirow,enumitem}
\usepackage[hidelinks]{hyperref}
\usepackage[capitalize,noabbrev]{cleveref}
\usepackage{color}

\usepackage{iftex}
\ifXeTeX
  \usepackage{fontspec}
\fi

\newcommand{\R}{\mathbb{R}}
\newcommand{\E}{\mathbb{E}}

\newcommand{\PP}{\mathbb{P}}
\newcommand{\GG}{\mathbb{G}}
\newcommand{\Normal}{\mathcal{N}}
\newcommand{\eps}{\varepsilon}

\newif\ifshowrev \showrevtrue
\AtBeginDocument{\ifshowrev\colorlet{revcolor}{blue}\else\colorlet{revcolor}{black}\fi}

\showrevfalse

\DeclareMathOperator{\tr}{tr}

\DeclareMathOperator{\CVaR}{CVaR}

\theoremstyle{plain}
\newtheorem{theorem}{Theorem}
\newtheorem{proposition}[theorem]{Proposition}
\newtheorem{lemma}[theorem]{Lemma}

\theoremstyle{definition}
\newtheorem{definition}{Definition}
\newtheorem{assumption}{Assumption}
\theoremstyle{remark}

\crefname{theorem}{Theorem}{Theorems}
\Crefname{theorem}{Theorem}{Theorems}
\crefname{proposition}{Proposition}{Propositions}
\Crefname{proposition}{Proposition}{Propositions}
\crefname{lemma}{Lemma}{Lemmas}
\Crefname{lemma}{Lemma}{Lemmas}
\crefname{corollary}{Corollary}{Corollaries}
\Crefname{corollary}{Corollary}{Corollaries}
\crefname{definition}{Definition}{Definitions}
\Crefname{definition}{Definition}{Definitions}
\crefname{assumption}{Assumption}{Assumptions}
\Crefname{assumption}{Assumption}{Assumptions}
\crefname{remark}{Remark}{Remarks}
\Crefname{remark}{Remark}{Remarks}

\makeatletter
\def\thm@space@setup{\thm@preskip=3\p@\@plus\p@\@minus\p@ \thm@postskip=\thm@preskip}
\patchcmd{\th@remark}{\thm@preskip\topsep \divide\thm@preskip\tw@}
  {\thm@preskip=3\p@\@plus\p@\@minus\p@}{}
  {\PackageError{preamble}{Patching \string\th@remark\space failed}{}}
\expandafter\patchcmd\csname\string\proof\endcsname
  {\topsep6\p@\@plus6\p@}{\topsep3\p@\@plus\p@\@minus\p@}{}
  {\PackageError{preamble}{Patching proof failed}{}}
\patchcmd{\thebibliography}{\section*{References}}
  {\@startsection{section}{1}{\z@}{1ex plus .5ex minus .5ex}{0.4ex}%
    {\normalfont\normalsize\centering\scshape}*{References}}{}
  {\PackageError{preamble}{Patching the References heading failed}{}}
\patchcmd{\thebibliography}
  {\vskip 0.3\baselineskip plus 0.1\baselineskip minus 0.1\baselineskip}{}{}
  {\PackageError{preamble}{Patching the bibliography skip failed}{}}
\patchcmd{\thebibliography}{\itemsep 0pt plus .5pt\relax}
  {\itemsep -1pt plus .5pt\relax\topsep\z@}{}
  {\PackageError{preamble}{Patching the bibliography item spacing failed}{}}
\makeatother
\setlist[itemize]{topsep=2pt,itemsep=0pt,parsep=0pt}

\title{\LARGE \bf Distributionally Robust Indirect-Feedback Stochastic MPC\\ with Gelbrich Ambiguity Sets}
\author{Jiaxun Sun, Riccardo Cescon$^\star$, Marcell Bartos$^\star$, Melanie N. Zeilinger%
\thanks{$^\star\,$Shared second author (equal contribution). This work has been submitted to the 2027 European Control Conference (ECC) for possible publication. {J.~Sun, M.~Bartos, and M.~N.~Zeilinger are with ETH Zurich, Switzerland (emails : \tt\small \{jiaxsun, mbartos, mzeilinger\}@ethz.ch\normalfont\small). R.~Cescon is with EPFL, Switzerland (emails : \tt\small riccardo.cescon@epfl.ch\normalfont\small). This work was supported as a part of NCCR Automation, a National Centre of Competence in Research, funded by the Swiss National Science Foundation (grant number 51NF40\textunderscore225155).}}%
}

\begin{document}
\maketitle
\thispagestyle{empty}
\pagestyle{empty}

\begin{abstract}
Stochastic model predictive control uses disturbance statistics to set safety margins. Errors in these statistics can invalidate the safety guarantees. This paper therefore describes the uncertainty in the disturbance mean and covariance by a prescribed Gelbrich ambiguity set. This contains distributions of any shape whose moments are close in the Gelbrich distance to the nominal values, including distributions with unbounded support. For linear time-invariant systems with additive disturbances, we build on the indirect feedback paradigm, propagate this moment uncertainty along the error dynamics, and obtain a closed-form constraint tightening that is computed offline. Robustness enters as a single additive term that grows with the size of the ambiguity set, so the online problem remains a quadratic program. Recursive feasibility and closed-loop satisfaction of conditional value-at-risk (CVaR) constraints hold for every disturbance distribution in the set. We illustrate the method on a linearized planar quadrotor.
\end{abstract}

\section{Introduction}\label{sec:intro}

Ensuring satisfaction of state and input constraints is challenging when the system is affected by unknown random disturbances. The usual way to address this is to use safety margins based on the mean and variance of the disturbance. Errors in these statistics can invalidate the risk guarantee the margins were meant to provide.

A standard approach in stochastic model predictive control (SMPC)~\cite{Mesbah2016,FarinaGiulioniScattolini2016} tightens constraints using the first two moments of the disturbance. We build on indirect-feedback stochastic MPC (IF-SMPC)~\cite{HewingZeilinger2020}, in which the nominal trajectory is never reset to the measurement. This separates the error dynamics from the online control decisions, allowing uncertainty bounds to be computed offline. The resulting guarantee, however, is stated for the assumed moments. Distributionally robust MPC removes that reliance by optimizing against a set of distributions~\cite{VanParysKuhnGoulartMorari2016,McAllisterEsfahani2025}.
In this paper, we combine IF-SMPC with a Gelbrich ambiguity set on the disturbance. The Gelbrich set bounds deviations in the mean and covariance without prescribing the shape of the distribution, so unbounded disturbances with finite second moments are admitted. A closed-form expression for the worst-case conditional value-at-risk (CVaR) over a Gelbrich set~\cite{NguyenShafieeFilipovicKuhn2023} turns the propagated moment uncertainty into a deterministic constraint tightening.

The main contributions of this paper are as follows.
\begin{itemize}
\item We construct an offline bound on how moment uncertainty accumulates over time and use it to determine constraint margins. Robustness enters as a single additive term that grows with the size of the ambiguity set, so the online problem remains a quadratic program.
\item We prove that, under the stated assumptions, the proposed controller is recursively feasible and satisfies each prescribed CVaR constraint, and hence the corresponding chance constraint, at every time step for every disturbance law in the ambiguity set.
\item We test the controller on a linearized planar quadrotor that flies into the corner between a wall and the floor, varying the disturbance moments and distribution shape. The baseline exceeds the risk limit in some tested cases, while the empirical CVaR of the proposed controller stays below it throughout.
\end{itemize}

\subsection{Related Work}\label{sec:related}

The deterministic margin used in this paper comes from the moment-based literature on chance and CVaR constraints~\cite{ElGhaouiOksOustry2003,CalafioreElGhaoui2006,ZymlerKuhnRustem2013}, with CVaR from~\cite{RockafellarUryasev2000}.

A related MPC approach with distributionally robust CVaR constraints is Wasserstein tube MPC~\cite{AolariteiFochesatoLygerosDorfler2023}, which propagates a Wasserstein tube exactly for noise with compact polyhedral support. Optimal-transport ambiguity-set propagation~\cite{AolariteiLanzettiChenDorfler2025} is the closest approach to the proposed offline propagation. An indirect-feedback scheme with Wasserstein DR-CVaR constraints~\cite{MarkLiu2021} restores recursive feasibility by softening the CVaR constraint. A later moment-based DR-MPC~\cite{MarkLiu2022} calibrates a radius on the second moment from data and restores feasibility by interpolating the initial state; in contrast, the proposed method uses a prescribed radius. Gelbrich ambiguity has also been used to robustify a quadratic cost under hard convex constraints~\cite{McAllisterEsfahani2025}, rather than a risk constraint.

Gelbrich ambiguity sets were introduced for Wasserstein distributionally robust Kalman filtering~\cite{ShafieezadehKuhn2018} and systematized in~\cite{NguyenShafieeFilipovicKuhn2023}. We use its closed-form expression for the worst-case CVaR. Distributionally robust density steering~\cite{PilipovskyTsiotras2024} uses the same closed form with a Gelbrich outer approximation over a finite horizon, optimizing feedback gains in a semidefinite program. In contrast, this paper propagates a Gelbrich ball offline along the indirect-feedback error dynamics, and the tightened nominal constraints need no slack variables.

\section{Preliminaries}\label{sec:prelim}

We consider a linear system with additive disturbances,
\begin{equation}\label{eq:system}
x(k+1) = A x(k) + B u(k) + w(k),
\end{equation}
with state $x(k)\in\R^n$, input $u(k)\in\R^m$, and i.i.d.\ disturbance $w(k)\in\R^n$. Throughout, $\|\cdot\|$ denotes the Euclidean norm and $\|\cdot\|_2$ the spectral norm. For symmetric matrices, $M\succeq 0$ ($M\succ 0$) means positive semidefinite (definite), and $M_1\succeq M_2$ means $M_1-M_2\succeq 0$. We write $\mathbb{S}_+^n$ for the symmetric positive semidefinite $n\times n$ matrices, $\PP^X$ for the law of $X$, and $A_\#\PP$ and $\PP*\mathbb{Q}$ for the laws of $AX$ and $X+Y$, where $X\sim\PP$ and $Y\sim\mathbb{Q}$ are independent.

\subsection{Gelbrich ambiguity sets}
\begin{definition}[Gelbrich distance~\cite{Gelbrich1990,KuhnShafieeWiesemann2025}]\label{def:gelbrich}
The Gelbrich distance between two mean--covariance pairs $(\mu_i,\Sigma_i)\in\R^n\times\mathbb{S}_+^n$, $i=1,2$, is defined as
\begin{equation}\label{eq:gelbrich-dist}
\begin{aligned}
&\mathrm{G}\bigl((\mu_1,\Sigma_1),(\mu_2,\Sigma_2)\bigr) := \Bigl(\|\mu_1-\mu_2\|^2 \\
&\quad + \tr\bigl(\Sigma_1+\Sigma_2-2(\Sigma_2^{1/2}\Sigma_1\Sigma_2^{1/2})^{1/2}\bigr)\Bigr)^{1/2}.
\end{aligned}
\end{equation}
\end{definition}

The Gelbrich ball of radius $\rho\ge 0$ centred at $(\hat\mu,\hat\Sigma)$ is
\begin{equation*}
\GG_\rho(\hat\mu,\hat\Sigma) := \bigl\{\PP\in\mathcal{P}_2(\R^n) : \mathrm{G}\bigl((\mu_\PP,\Sigma_\PP),(\hat\mu,\hat\Sigma)\bigr)\le\rho\bigr\},
\end{equation*}
where $\mathcal{P}_2(\R^n)$ is the set of distributions over $\mathbb{R}^n$ with finite second moments, and $\mu_\PP$ and $\Sigma_\PP$ are the mean and covariance of $\PP$. The true disturbance law is unknown and is assumed to satisfy $\PP^w\in\GG_{\rho_w}(\hat\mu_w,\hat\Sigma_w)$ with $\hat\Sigma_w\succ 0$. That is, the mean--covariance pair of the true disturbance lies within Gelbrich distance $\rho_w$ of the nominal pair $(\hat\mu_w,\hat\Sigma_w)$. Membership depends only on the mean and covariance, so the set also contains laws with unbounded support, such as $\Normal(\hat\mu_w,\hat\Sigma_w)$.\looseness=-1

\begin{assumption}\label{ass:zero-nominal-mean}
The nominal disturbance mean satisfies $\hat\mu_w = 0$. The true disturbance mean may be nonzero, provided that the disturbance law lies in the ambiguity set.
\end{assumption}

\subsection{Risk constraints}
State and input are subject to the half-space constraints
\begin{equation}\label{eq:halfspace}
\begin{aligned}
\mathcal{X}^j &= \{x : h_j^\top x \le b_j\}, && j=1,\dots,n_x, \\
\mathcal{U}^l &= \{u : g_l^\top u \le c_l\}, && l=1,\dots,n_u.
\end{aligned}
\end{equation}
Here, $h_j\in\R^n$, $g_l\in\R^m$, and $b_j,c_l\in\R$, and $n_x\in\mathbb{N}$ and $n_u\in\mathbb{N}$ denote the numbers of state and input constraints. Under stochastic disturbances, these constraints are imposed as chance constraints $\PP(h_j^\top x(k)\le b_j)\ge 1-\eps^x_j$ and $\PP(g_l^\top u(k)\le c_l)\ge 1-\eps^u_l$, with $\eps^x_j,\eps^u_l\in(0,1)$. Chance constraints need not be convex~\cite{NemirovskiShapiro2006}, we therefore use CVaR constraints as a conservative approximation.

For a tail probability $\eps\in(0,1)$, the conditional value-at-risk of a random loss $Y$ is
\begin{equation*}
\CVaR_\eps(Y)
:= \inf_{\tau\in\R}\bigl\{\tau + \eps^{-1}\E[(Y-\tau)_+]\bigr\}.
\end{equation*}
For continuous distributions, it is the expected value of $Y$ over its worst $\eps$-fraction of outcomes. For state constraints, the loss is $Y=h_j^\top x(k)-b_j$, whose positive values are violations. Relevant properties of CVaR are that 1) the inequality $\CVaR_\eps(Y)\le 0$ implies $\PP(Y>0)\le\eps$ \cite{RockafellarUryasev2000}, i.e., a CVaR constraint is a conservative approximation of a chance constraint and also accounts for the magnitude of violations; 2) for a fixed uncertainty distribution, a CVaR constraint is convex when the loss is convex in the decision variables~\cite{RockafellarUryasev2000}; and 3) CVaR is monotone and satisfies $\CVaR_\eps(Y+c)=\CVaR_\eps(Y)+c$ for every deterministic constant $c$ \cite{RockafellarUryasev2002}.

The design goal is that each state and input constraint in~\eqref{eq:halfspace} satisfies a distributionally robust CVaR constraint at every time. For every $k\ge 0$, every $j$, and every $l$,
\begin{equation}\label{eq:dr-cvar-goal}
\begin{aligned}
\sup_{\PP^w\in\GG_{\rho_w}(\hat\mu_w,\hat\Sigma_w)}\ \CVaR_{\eps^x_j}\bigl(h_j^\top x(k)-b_j\bigr) &\le 0, \\
\sup_{\PP^w\in\GG_{\rho_w}(\hat\mu_w,\hat\Sigma_w)}\ \CVaR_{\eps^u_l}\bigl(g_l^\top u(k)-c_l\bigr) &\le 0\,.
\end{aligned}
\end{equation}
Each bound applies to one constraint at one time step, under the unconditional closed-loop law induced by $\PP^w$.
\begin{lemma}[Worst-case CVaR over a Gelbrich ball {\cite[Thm.~5, Prop.~8(i)]{NguyenShafieeFilipovicKuhn2023}}]\label{lem:wc-cvar}
Let $\PP$ be the law of a random vector $X\in\R^n$, and evaluate CVaR under $\PP$. For every $h\in\R^n$, $b\in\R$, $\rho\ge 0$, and $\eps\in(0,1)$,
\begin{equation*}
\begin{aligned}
&\sup_{\PP\in\GG_\rho(\hat\mu,\hat\Sigma)}\CVaR_\eps\bigl(h^\top X-b\bigr) \\
&\qquad = h^\top\hat\mu - b + \kappa(\eps)\sqrt{h^\top\hat\Sigma h} + \frac{\rho}{\sqrt{\eps}}\|h\|,
\end{aligned}
\end{equation*}
where $\kappa(\eps):=\sqrt{(1-\eps)/\eps}$.
\end{lemma}

\section{Methodology}\label{sec:method}

IF-SMPC~\cite{HewingZeilinger2020} tightens the constraints on a nominal trajectory by time-varying margins computed offline. This paper computes these margins for every disturbance law in the Gelbrich ball, not only for the nominal moments.

\subsection{Indirect-feedback decomposition}
As in IF-SMPC~\cite{HewingZeilinger2020}, we split the state and input into nominal and error parts, with stabilizing gain $K$,
\begin{equation}\label{eq:tube}
x(k) = z(k) + e(k), \qquad u(k) = v(k) + K e(k).
\end{equation}
Substituting~\eqref{eq:tube} into~\eqref{eq:system} gives the nominal and error dynamics
\begin{align}
z(k+1) &= A z(k) + B v(k), \label{eq:nominal}\\
e(k+1) &= A_K e(k) + w(k), \label{eq:error}
\end{align}
with $A_K := A+BK$, which is Schur stable. The nominal state $z$ is never reset to the measured state. We adopt the initialization
\begin{equation}\label{eq:init}
z(0) = x(0), \qquad\text{equivalently}\qquad e(0) = 0.
\end{equation}

\subsection{Offline error envelope}
The error dynamics~\eqref{eq:error} do not depend on the online input $v(k)$, so we enclose the law of $e(k)$ in a Gelbrich ball offline in \cref{prop:propagation}. \Cref{lem:linear-map,lem:indep-sum} give the two properties of Gelbrich balls needed for this. Both use that the Gelbrich distance between two mean--covariance pairs equals the type-2 Wasserstein distance $\mathrm{W}_2$, the infimum of $(\E\|X-Y\|^2)^{1/2}$ over all couplings, between the Gaussians with these moments~\cite{Gelbrich1990}. The lemmas follow from the standard propagation bounds for Wasserstein balls under linear maps and convolutions~\cite[Thms.~6 and~11]{AolariteiLanzettiChenDorfler2025} combined with this Gaussian formula; we give short proofs for completeness.

\begin{lemma}\label{lem:linear-map}
For every $A\in\R^{n\times n}$,
\begin{equation*}
A_\#\GG_\rho(\hat\mu,\hat\Sigma)
\subseteq \GG_{\|A\|_2\rho}\bigl(A\hat\mu,\, A\hat\Sigma A^\top\bigr).
\end{equation*}
\end{lemma}
\begin{proof}
Let $\PP\in\GG_\rho(\hat\mu,\hat\Sigma)$ have mean $\mu$ and covariance $\Sigma$, and introduce the moment-matched Gaussians $\nu:=\Normal(\mu,\Sigma)$ and $\hat\nu:=\Normal(\hat\mu,\hat\Sigma)$; $\PP$ itself need not be Gaussian. Then $\mathrm{W}_2(\nu,\hat\nu)=\mathrm{G}((\mu,\Sigma),(\hat\mu,\hat\Sigma))\le\rho$. The law $A_\#\PP$ has mean $A\mu$ and covariance $A\Sigma A^\top$, the moments of the Gaussian $A_\#\nu$, and $A_\#\hat\nu=\Normal(A\hat\mu,A\hat\Sigma A^\top)$. Mapping a coupling of $\nu$ and $\hat\nu$ by $(x,y)\mapsto(Ax,Ay)$ gives $\mathrm{W}_2(A_\#\nu,A_\#\hat\nu)\le\|A\|_2\,\mathrm{W}_2(\nu,\hat\nu)$, cf.~\cite[Thm.~6]{AolariteiLanzettiChenDorfler2025}. Hence
\begin{equation*}
\begin{aligned}
&\mathrm{G}\bigl((A\mu,A\Sigma A^\top),(A\hat\mu,A\hat\Sigma A^\top)\bigr)\\
&\quad=\mathrm{W}_2(A_\#\nu,A_\#\hat\nu)\le\|A\|_2\,\rho,
\end{aligned}
\end{equation*}
that is, $A_\#\PP\in\GG_{\|A\|_2\rho}(A\hat\mu,A\hat\Sigma A^\top)$.
\end{proof}

\begin{lemma}\label{lem:indep-sum}
Let $X$ and $Y$ be independent with $\PP^X\in\GG_{\rho_X}(\hat\mu_X,\hat\Sigma_X)$ and $\PP^Y\in\GG_{\rho_Y}(\hat\mu_Y,\hat\Sigma_Y)$. Then
\begin{equation*}
\PP^{X+Y}\in\GG_{\rho_X+\rho_Y}\bigl(\hat\mu_X+\hat\mu_Y,\,\hat\Sigma_X+\hat\Sigma_Y\bigr).
\end{equation*}
\end{lemma}
\begin{proof}
Let $\PP^X$ and $\PP^Y$ have moments $(\mu_X,\Sigma_X)$ and $(\mu_Y,\Sigma_Y)$, and introduce the moment-matched Gaussians $\nu_X:=\Normal(\mu_X,\Sigma_X)$ and $\hat\nu_X:=\Normal(\hat\mu_X,\hat\Sigma_X)$, and $\nu_Y$ and $\hat\nu_Y$ likewise. By independence, $X+Y$ has mean $\mu_X+\mu_Y$ and covariance $\Sigma_X+\Sigma_Y$, the moments of the Gaussian $\nu_X*\nu_Y$, and $\hat\nu_X*\hat\nu_Y=\Normal(\hat\mu_X+\hat\mu_Y,\hat\Sigma_X+\hat\Sigma_Y)$. The convolution bound~\cite[Thm.~11]{AolariteiLanzettiChenDorfler2025} for the type-2 Wasserstein distance gives
\begin{equation*}
\begin{aligned}
&\mathrm{W}_2(\nu_X*\nu_Y,\hat\nu_X*\hat\nu_Y)\\
&\quad\le\mathrm{W}_2(\nu_X,\hat\nu_X)+\mathrm{W}_2(\nu_Y,\hat\nu_Y)\le\rho_X+\rho_Y.
\end{aligned}
\end{equation*}
By the Gaussian formula, the left-hand side is the Gelbrich distance from the moments of $X+Y$ to the claimed centre.
\end{proof}

Applying \cref{lem:linear-map,lem:indep-sum} to the error dynamics~\eqref{eq:error} yields the following recursions for the centre covariance and radius, starting from $\hat\Sigma_e(0)=0$ and $\rho_e(0)=0$:
\begin{align}
\hat\Sigma_e(t+1) &= A_K \hat\Sigma_e(t) A_K^\top + \hat\Sigma_w, \label{eq:cov-rec}\\
\rho_e(t+1) &= \rho_e(t) + \|A_K^t\|_2\, \rho_w. \label{eq:rho-rec}
\end{align}

\begin{proposition}\label{prop:propagation}
Under \cref{ass:zero-nominal-mean} and~\eqref{eq:init}, every $\PP^w\in\GG_{\rho_w}(\hat\mu_w,\hat\Sigma_w)$ satisfies
\begin{equation}\label{eq:membership}
\PP^{e(k)} \in \GG_{\rho_e(k)}\bigl(0,\,\hat\Sigma_e(k)\bigr), \qquad \forall k \geq  0.
\end{equation}
\end{proposition}
\begin{proof}
The claim holds for $k=0$ because $e(0)=0$. For $k\ge 1$, \eqref{eq:error} and~\eqref{eq:init} give $e(k)=\sum_{i=0}^{k-1}A_K^i w(k-1-i)$. By \cref{ass:zero-nominal-mean} and \cref{lem:linear-map}, the law of the term $A_K^i w(k-1-i)$ lies in $\GG_{\|A_K^i\|_2\rho_w}(0,A_K^i\hat\Sigma_w(A_K^i)^\top)$. The terms are independent because the disturbances are i.i.d. Applying \cref{lem:indep-sum} repeatedly, the law of $e(k)$ lies in the Gelbrich ball with centre $\bigl(0,\sum_{i=0}^{k-1}A_K^i\hat\Sigma_w(A_K^i)^\top\bigr)$ and radius $\rho_w\sum_{i=0}^{k-1}\|A_K^i\|_2$. By~\eqref{eq:cov-rec} and~\eqref{eq:rho-rec}, the covariance sum is $\hat\Sigma_e(k)$ and the radius is $\rho_e(k)$.
\end{proof}

The terminal set in \cref{sec:mpc} requires the limits of these sequences. The sequences converge to the steady-state values
\begin{equation}\label{eq:steady}
\hat\Sigma_e^\infty = A_K \hat\Sigma_e^\infty A_K^\top + \hat\Sigma_w, \qquad
\rho_e^\infty = \rho_w\sum_{i=0}^{\infty}\|A_K^i\|_2,
\end{equation}
where the series converges because $A_K$ is Schur stable. The sequences are monotone: $\hat\Sigma_e(t)\preceq\hat\Sigma_e(t+1)\preceq\hat\Sigma_e^\infty$ and $\rho_e(t)\le\rho_e(t+1)\le\rho_e^\infty$. For computation, we choose an integer $p\ge 1$ with $\gamma_p:=\|A_K^p\|_2<1$, which exists since $A_K^p\to 0$, and use the upper bound
\begin{equation}\label{eq:rho-bar}
\rho_e^\infty \le \bar\rho_e^\infty := \frac{\rho_w}{1-\gamma_p}\sum_{i=0}^{p-1}\|A_K^i\|_2.
\end{equation}
Writing $i=qp+r$ with $0\le r<p$, submultiplicativity gives $\|A_K^i\|_2\le\gamma_p^q\|A_K^r\|_2$; summing over $q$ and $r$ gives the bound.

\subsection{Constraint tightening}
We derive sufficient conditions for the DR-CVaR constraints~\eqref{eq:dr-cvar-goal}, which allow the constraint tightening to be computed offline. By~\eqref{eq:tube}, the error terms in the state and input losses are $h_j^\top e(k)$ and $(K^\top g_l)^\top e(k)$. Applying \cref{lem:wc-cvar} with the ambiguity set in~\eqref{eq:membership} gives the tightened bounds
\begin{align}
\tilde b_j(t)
&:= b_j - \kappa(\eps^x_j)\sqrt{h_j^\top\hat\Sigma_e(t)h_j}
- \frac{\rho_e(t)}{\sqrt{\eps^x_j}}\|h_j\|, \label{eq:tilde-b}\\
\tilde c_l(t)
&:= c_l - \kappa(\eps^u_l)\sqrt{g_l^\top K\hat\Sigma_e(t)K^\top g_l}
- \frac{\rho_e(t)}{\sqrt{\eps^u_l}}\|K^\top g_l\|. \label{eq:tilde-c}
\end{align}
For $\rho_w=0$, the last terms vanish and the bounds reduce to the moment-based (Chebyshev--Cantelli) tightening of IF-SMPC~\cite{HewingZeilinger2020,ZymlerKuhnRustem2013}. The nominal state and input may be random, so these bounds are imposed along every sample path. If $h_j^\top z(k)\le\tilde b_j(k)$ holds pathwise, then
\begin{equation}\label{eq:pathwise}
h_j^\top x(k) - b_j
\le h_j^\top e(k) + \bigl(\tilde b_j(k) - b_j\bigr).
\end{equation}
The shift $\tilde b_j(k)-b_j$ is deterministic. By \cref{prop:propagation} and \cref{lem:wc-cvar}, the CVaR of $h_j^\top e(k)$ is at most $b_j-\tilde b_j(k)$. Monotonicity and the translation property of CVaR therefore give $\CVaR_{\eps^x_j}(h_j^\top x(k)-b_j)\le 0$. The same argument applies to the input constraints. These constraints define the tightened polyhedra
\begin{equation*}
\begin{aligned}
\tilde{\mathcal{X}}(t) &:= \{z : h_j^\top z \le \tilde b_j(t),\,\forall j\}, \\
\tilde{\mathcal{U}}(t) &:= \{v : g_l^\top v \le \tilde c_l(t),\,\forall l\}.
\end{aligned}
\end{equation*}
Since $\tilde b_j(t)$ and $\tilde c_l(t)$ are nonincreasing in $t$, the tightened sets are nested:
\begin{equation}\label{eq:nested}
\begin{aligned}
\tilde{\mathcal{X}}(0)&\supseteq\tilde{\mathcal{X}}(1)\supseteq\cdots\supseteq\tilde{\mathcal{X}}(\infty), \\
\tilde{\mathcal{U}}(0)&\supseteq\tilde{\mathcal{U}}(1)\supseteq\cdots\supseteq\tilde{\mathcal{U}}(\infty),
\end{aligned}
\end{equation}
where the sets at infinity use $\hat\Sigma_e^\infty$ from~\eqref{eq:steady} and $\bar\rho_e^\infty$ from~\eqref{eq:rho-bar}. The sets at infinity are conservative subsets of the limiting tightened sets.

\subsection{MPC formulation}\label{sec:mpc}
Consider prediction horizon length $N\ge 1$ and stage costs $(Q,R)\succ 0$, and let the terminal cost weight $P\succ 0$ solve $A_K^\top P A_K - P = -(Q+K^\top R K)$.
At time $k$ the measured error $e(k)=x(k)-z(k)$ is known. Under the nominal zero-mean disturbance model, its predicted mean is $\bar e_i := A_K^i e(k)$. The measured error enters only the objective, through $\bar e_i$, while the nominal initial state remains $z(k)$. After omitting terms independent of the decision variables, the conditional expected quadratic cost under this model reduces to
\begin{equation}\label{eq:mpc}
\begin{aligned}
\min_{v_0,\dots,v_{N-1}} \
& \sum_{i=0}^{N-1} \bigl(\|z_i+\bar e_i\|_Q^2 + \|v_i+K\bar e_i\|_R^2\bigr) + \|z_N+\bar e_N\|_P^2 \\
\text{s.t.} \
& z_0 = z(k), \\
& z_{i+1} = A z_i + B v_i, \ i=0,\dots,N-1 \\
& z_i\in\tilde{\mathcal{X}}(k+i),\ 
  v_i\in\tilde{\mathcal{U}}(k+i), \ i=0,\dots,N-1 \\
& z_N\in\mathcal{X}_f,
\end{aligned}
\end{equation}
where $\|y\|_M^2:=y^\top My$.
At each time $k$, we solve~\eqref{eq:mpc}, apply $u(k)=v_0^\star(k)+K e(k)$, where $v_0^\star(k)$ is the first element of the optimal input sequence, and propagate the nominal state as $z(k+1)=A z(k)+B v_0^\star(k)$. Because the margins are indexed by the absolute time $k+i$, they follow the offline envelope rather than being reset at each solve.

The terminal set $\mathcal{X}_f$ must satisfy the following standard assumption, cf.~\cite{HewingZeilinger2020}.
\begin{assumption}\label{ass:terminal}
The terminal set $\mathcal{X}_f$ is nonempty, polyhedral, positively invariant under $A_K$, and satisfies
\begin{equation}\label{eq:terminal}
\mathcal{X}_f \subseteq \tilde{\mathcal{X}}(\infty)\cap\{z : Kz\in\tilde{\mathcal{U}}(\infty)\}.
\end{equation}
\end{assumption}

\section{Theoretical Guarantees}\label{sec:results}

We now state and prove the theoretical guarantees of the proposed method in terms of recursive feasibility and closed-loop constraint satisfaction.

\begin{theorem}[Recursive feasibility]\label{thm:rf}
Let \cref{ass:terminal} hold. If~\eqref{eq:mpc} is feasible at $k=0$, then it is feasible at every $k\ge 0$ for every disturbance realization.
\end{theorem}
\begin{proof}
Let $(v_0^\star,\dots,v_{N-1}^\star)$ be feasible at time $k$, with nominal trajectory $(z_0^\star,\dots,z_N^\star)$. The nominal update gives $z(k+1)=z_1^\star$ for every disturbance realization, because~\eqref{eq:nominal} contains no disturbance. At time $k+1$, consider the shifted candidates $(v_1^\star,\dots,v_{N-1}^\star,Kz_N^\star)$ and $(z_1^\star,\dots,z_N^\star,A_Kz_N^\star)$. That is, for prediction steps $i=0,\dots,N-2$, the candidate pairs are the pairs of the time-$k$ solution at step $i+1$. Both refer to the same absolute time $k+1+i$, so they satisfy the same constraints. For $i=N-1$, $z_N^\star\in\mathcal{X}_f$, so~\eqref{eq:terminal} and~\eqref{eq:nested} give $z_N^\star\in\tilde{\mathcal{X}}(\infty)\subseteq\tilde{\mathcal{X}}(k+N)$ and $Kz_N^\star\in\tilde{\mathcal{U}}(\infty)\subseteq\tilde{\mathcal{U}}(k+N)$. Finally, $A_Kz_N^\star\in\mathcal{X}_f$ by positive invariance. The candidate is therefore feasible at time $k+1$, and the claim follows by induction.
\end{proof}
Pathwise feasibility then yields closed-loop CVaR constraint satisfaction.

\begin{theorem}[Closed-loop CVaR constraint satisfaction]\label{thm:cvar}
Let \cref{ass:zero-nominal-mean,ass:terminal} and the initialization~\eqref{eq:init} hold. If~\eqref{eq:mpc} is feasible at $k=0$, then the closed loop under $u(k)=v_0^\star(k)+Ke(k)$ satisfies~\eqref{eq:dr-cvar-goal} for every $k\ge 0$, every $j$, and every $l$.
\end{theorem}
\begin{proof}
Fix $k$, $j$, and $\PP^w\in\GG_{\rho_w}(\hat\mu_w,\hat\Sigma_w)$. By \cref{thm:rf}, \eqref{eq:mpc} is feasible at time $k$. Its constraint at $i=0$ gives $h_j^\top z(k)\le\tilde b_j(k)$ along every sample path, so~\eqref{eq:pathwise} holds. The argument after~\eqref{eq:pathwise}, which uses \cref{prop:propagation} and does not need $z(k)$ and $e(k)$ to be independent, then gives $\CVaR_{\eps^x_j}(h_j^\top x(k)-b_j)\le 0$. Since $\PP^w$ was arbitrary, the state inequality in~\eqref{eq:dr-cvar-goal} holds. For the inputs, $v_0^\star(k)\in\tilde{\mathcal{U}}(k)$ gives $g_l^\top u(k)-c_l\le(K^\top g_l)^\top e(k)+\tilde c_l(k)-c_l$ pathwise, and the same argument with~\eqref{eq:tilde-c} gives the input inequality.
\end{proof}

By property~1) of CVaR, \cref{thm:cvar} also implies satisfaction of the corresponding chance constraints. Together, the two theorems extend the guarantees of IF-SMPC~\cite{HewingZeilinger2020} from the nominal moments to every disturbance law in the Gelbrich ball. The price is the additive term $\rho_e(t)\|h_j\|/\sqrt{\eps^x_j}$ in the tightening, which grows linearly with $\rho_w$. Since the Gelbrich distance lower-bounds the type-2 Wasserstein distance~\cite{Gelbrich1990}, the Gelbrich ball contains the type-2 Wasserstein ball of equal radius around any law with moments $(\hat\mu_w,\hat\Sigma_w)$ in the same coordinates. This larger set covers more laws but can lead to more conservative tightening. The guarantees, however, rely on i.i.d. disturbances: independence allows the radii to be added in \cref{lem:indep-sum}, so disturbances that are correlated over time are not covered, unlike in the case of IF-SMPC.\looseness=-1

\section{Numerical Results}\label{sec:numerics}

We test the proposed controller, DR-MPC, on a linearized planar quadrotor. It flies into the corner between a wall and the floor while a steady wind pushes it toward both. Code and data are available at \url{https://github.com/JinnoNian/Gelbrich-DR-IFSMPC-numerics}.

\subsection{System and constraints}
The state is $x=(p_x,p_z,\theta,v_x,v_z,\omega)$: horizontal and vertical position, pitch angle, and their rates. The input $u=(\Delta F,\tau)$ is the thrust deviation from hover and the torque. The model is the planar hover linearization of the quadrotor model in \cite{MahonyKumarCorke2012} with mass $1$\,kg and inertia $0.02$\,kg\,m$^2$, discretized by zero-order hold with $T_s=0.1$\,s.
The origin lies 1\,m in front of the wall and 0.6\,m above the floor, with $p_x$ pointing toward the wall and $p_z$ pointing up. The quadrotor starts at rest at $x(0)=(-2.5,1,0,0,0,0)$ and should reach the corner $x_{\mathrm{ref}}=(1,-0.6,0,0,0,0)$. The state constraints are $-4.5\le p_x\le1$ (wall), $-0.6\le p_z\le3$ (floor), $|\theta|\le0.35$, $|v_x|,|v_z|\le1.5$, and $|\omega|\le2$. The input constraints are $|\Delta F|\le4.905$ and $|\tau|\le0.2$. This gives $n_x=12$ and $n_u=4$ half-spaces, which are all imposed as CVaR constraints with $\eps^x_j=\eps^u_l=\eps=0.2$ for all $j,l$.\looseness=-1

\subsection{Disturbance model and controller}
The nominal disturbance law has mean zero and diagonal covariance $\hat\Sigma_w$. Its standard deviations are 2\,mm on $p_x,p_z$, 0.002\,rad on $\theta$, 0.03\,m/s on $v_x,v_z$, and 0.02\,rad/s on $\omega$. We tuned these values to obtain centimetre-scale nominal steady-state position standard deviations (4.8\,cm in $p_x$ and 2.2\,cm in $p_z$ under the nominal covariance) and visible tightening margins, while retaining feasibility for the chosen setup. The velocity and angular-rate disturbance standard deviations correspond in scale to forces of 0.3\,N per translational direction (3\% of the hover thrust) and a torque of 0.004\,N\,m (2\% of the torque limit) over one sampling period. The Gelbrich distance and matrix norms depend on units, so all controller computations use scaled coordinates $S_x^{-1}x$ and $S_u^{-1}u$ with $S_x=\mathrm{diag}(1,1,0.35,1,1,2)$ and $S_u=\mathrm{diag}(4.905,0.2)$. The quantities $\rho_w$, $Q$, $R$, and $K$ below refer to these coordinates. The radius is $\rho_w=0.006$, about 14\% of $\sqrt{\tr\hat\Sigma_w}$.\looseness=-1

We use $N=20$, $Q=\mathrm{diag}(10,10,1,1,1,1)$, and $R=I$. The reference lies on the wall and floor boundaries to make the effect of constraint tightening visible. The state terms in~\eqref{eq:mpc} penalize $z_i+\bar e_i-x_{\mathrm{ref}}$. This only changes the cost, so \cref{thm:rf,thm:cvar} still apply. The gain $K$ is the LQR gain for the weights $(I,3I)$. With $p=371$, the bound~\eqref{eq:rho-bar} gives $\bar\rho_e^\infty=0.160\approx27\rho_w$. For each controller, $\mathcal{X}_f$ is the maximal positively invariant subset of the right-hand side of~\eqref{eq:terminal}~\cite{GilbertTan1991}. We solve~\eqref{eq:mpc} with the interior-point solver Clarabel~\cite{GoulartChen2024}.

\subsection{Compared controllers}
All controllers share the objective, dynamics, $K$, horizon, and solver. They differ only in the constraint tightening and in the terminal set derived from it.\looseness=-1
\begin{itemize}[leftmargin=*]
\item DR-MPC: the proposed controller with $\rho_w=0.006$.
\item IF-SMPC~\cite{HewingZeilinger2020}: the same controller with $\rho_w=0$.
\item COV: IF-SMPC with the inflated covariance $(1+\rho_w/\sqrt{\tr\hat\Sigma_w})^2\hat\Sigma_w$, the largest scaling of $\hat\Sigma_w$ inside the Gelbrich ball. It has no mean term.
\item ORC: an oracle that applies the tightening of IF-SMPC with the true mean and covariance of each scenario. It uses the running maximum of each true-moment margin; a tail bound defines the terminal tightening.
\item W2: the type-2 Wasserstein ball of radius $\rho_w$ around $\Normal(0,\hat\Sigma_w)$ replaces the Gelbrich ball. By \cref{sec:results}, this ball lies in the Gelbrich ball, and it has the same Gaussian members~\cite{Gelbrich1990}. With the radii $\rho_e(k)$ of DR-MPC, \cite[Prop.~6.20]{KuhnShafieeWiesemann2025} gives the margin of DR-MPC with $\kappa(\eps)=2$ replaced by $\varphi(\Phi^{-1}(1-\eps))/\eps=1.40$, where $\varphi$ and $\Phi$ are the standard normal pdf and cdf. With its own terminal set, the argument of \cref{thm:rf} gives recursive feasibility if the initial problem is feasible, but we prove no closed-loop CVaR bound for W2; it is a numerical baseline at the same radius.\looseness=-1
\end{itemize}

\subsection{True disturbances and evaluation}
We test robustness to misspecification of the disturbance moments
and distribution shape. The true disturbance is
$w=\mu+\Sigma^{1/2}\xi$, where $\xi$ is i.i.d.\ over time,
with zero mean and identity covariance. We parameterize the
true moments as
\begin{equation*}
\mu=\sqrt{\eta}\,\delta\rho_w\,d,\quad \Sigma=a^2\hat\Sigma_w,\quad a=1+\sqrt{1-\eta}\,\frac{\delta\rho_w}{\sqrt{\tr\hat\Sigma_w}}.
\end{equation*}
where $d$ is a unit vector, $\delta$ controls the magnitude
of misspecification, and $\eta$ determines how it is split
between the mean and covariance. By construction, the
Gelbrich distance between the true and nominal moments is
$\delta\rho_w$. We use $\eta=0.75$ and
$\delta\in\{0,0.5,1,1.5\}$, so the true moments lie within
the DR-MPC ambiguity set for $\delta\le1$ and outside it
for $\delta=1.5$. We choose
$d\approx(0,0,0,0.911,-0.413,0)$ to introduce a
wind-like bias in $(v_x,v_z)$, producing a steady-state
error mean $(I-A_K)^{-1}\mu$ toward the wall and floor.

For each $\delta$, we consider four shapes for $\xi$,
which do not change the disturbance moments: Gaussian, standardized
multivariate Student-$t$ with 5 degrees of freedom,
multivariate Laplace $\xi=\sqrt{\lambda}\,\zeta$, and a
bimodal Gaussian mixture $\xi=s\alpha d+\zeta'$. Here,
$\lambda\sim\mathrm{Exp}(1)$, $\zeta\sim\Normal(0,I)$,
$s\in\{-1,1\}$ with equal probabilities, $\alpha=0.85$,
and $\zeta'\sim\Normal(0,I-\alpha^2dd^\top)$ are independent.
For $\Sigma=a^2\hat\Sigma_w$, the two distances agree
only if $w$ is Gaussian, so at $\delta=1$ only the Gaussian
law lies in the W2 ball; for $\delta<1$, the moments
leave it open.

For each combination of $\delta$ and distribution shape,
we simulate 2000 trajectories of 80 steps, using the
same disturbance samples for all controllers.
At each time step, we estimate the CVaR of the
normalized constraint loss
$(h_j^\top x(k)-b_j)/\|h_j\|$
for every state constraint, and analogously for the
input constraints. We define the worst risk as the
maximum estimated CVaR over all constraints and
times $k\ge1$.\looseness=-1

We quantify estimation uncertainty with a trajectory-level
bootstrap~\cite{EfronTibshirani1993}. For each scenario,
we resample the 2000 trajectories with replacement 1000
times, with identical indices across controllers. In each
resample, we recompute all CVaR estimates and record their
largest absolute deviation from the original estimates
over all controllers, constraints, and times. Adding
and subtracting the 95th percentile of these deviations as
a common margin gives approximate simultaneous
95\% confidence intervals within each scenario,
including for the worst risk.\looseness=-1

\begin{figure*}[t]
\centering
\includegraphics[width=\textwidth,trim=0 4 0 5.5,clip]{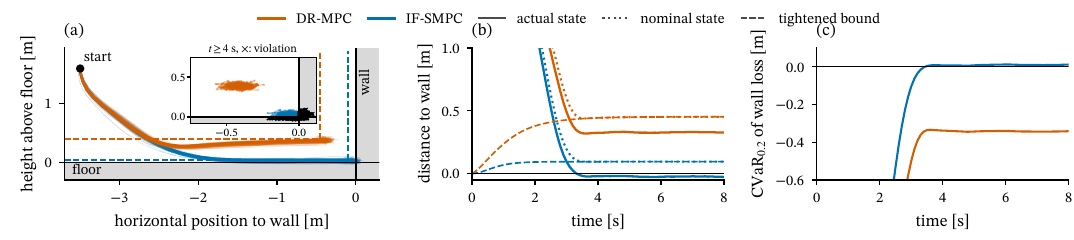}
\caption{Closed-loop system under Student-$t$ disturbances with $\delta=1$. (a) First 100 flight paths, their medians, and the steady-state tightened bounds (dashed). Inset: positions after 4\,s, crosses beyond the wall or the floor. (b) Distance to the wall: 5\% quantile of the actual state (solid), median of the nominal state (dotted), and margin $b-\tilde b(k)$ (dashed). (c) Empirical CVaR of the wall loss. Pointwise 95\% intervals (widths below 0.01\,m) are omitted.}
\label{fig:closed-loop}
\end{figure*}

\subsection{Effect of the tightening}
\Cref{fig:closed-loop} shows the Student-$t$ case with $\delta=1$, on the boundary of the ambiguity set. DR-MPC and IF-SMPC fly to the corner and keep a distance set by the margin $b-\tilde b(k)$. In steady state, the wall margin is 0.45\,m for DR-MPC and 0.10\,m for IF-SMPC; the difference is the robust term $\rho_e(k)\|h_j\|/\sqrt{\eps^x_j}$. The wind pushes the actual state toward the wall. After 4\,s, about 12\% of the IF-SMPC trajectories are beyond the wall at each time, and its wall CVaR settles near $+0.01$\,m. No DR-MPC trajectory crosses the wall after 4\,s, and its wall CVaR stays near $-0.34$\,m. The price is a closed-loop cost 17--18\% above IF-SMPC in every scenario.

\subsection{Robustness to misspecification}
\begin{figure}[t]
\centering
\includegraphics[width=\columnwidth,trim=0 0 0 6,clip]{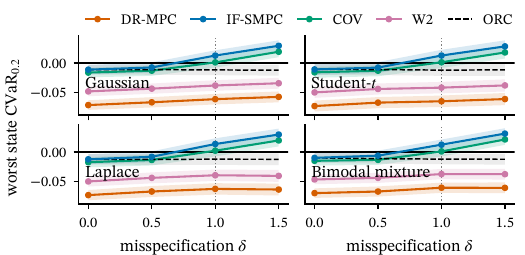}
\caption{Worst state CVaR over constraints and times versus the misspecification level $\delta$, with approximate 95\% intervals.}
\label{fig:misspec}
\end{figure}

\Cref{fig:misspec} shows all scenarios, and \cref{tab:summary} lists the scenario of \cref{fig:closed-loop}. The four distribution shapes give similar results, so in these tests the moments matter more than the shape. DR-MPC keeps every interval below zero, also at $\delta=1.5$, so the worst case over the ball is conservative for these distributions. Its worst constraint is the lower bound on $v_z$ early in the descent. IF-SMPC meets the risk limit at $\delta=0$, as its guarantee predicts. Its intervals contain zero at $\delta=0.5$ and lie above zero for $\delta\ge1$, with the wall as the worst constraint. All controllers satisfy the input CVaR constraints.\looseness=-1

COV inflates only the covariance. It lowers the risk of IF-SMPC, but its intervals contain zero at $\delta=1$ and lie above zero at $\delta=1.5$. A larger variance margin thus does not replace a margin for the mean shift. ORC meets the risk limit at every $\delta$, at a cost less than 2\% above IF-SMPC. It shows what the risk limit costs when the true moments are known, whereas DR-MPC covers every law in the ball. In the Gaussian scenarios with $\delta\le1$, which lie in both balls, W2 keeps its intervals below zero with worst risk $-0.048$ to $-0.038$ (DR-MPC: $-0.072$ to $-0.061$) at a cost 14.7--15.5\% above IF-SMPC. W2 also keeps its intervals below zero for laws outside its ball, such as the Student-$t$ law of \cref{tab:summary}. Bounding the whole distribution thus costs less in these tests, but the W2 ball covers fewer disturbance laws.\looseness=-1

\begin{table}[t]
\caption{Scenario of \cref{fig:closed-loop}: worst state CVaR (simultaneous 95\% intervals), largest violation frequency, cost relative to IF-SMPC, median time per step.}
\label{tab:summary}
\centering
\setlength{\tabcolsep}{3pt}
\begin{tabular}{lcccc}
\toprule
 & Worst state CVaR & Viol.\ [\%] & Cost [\%] & Time [ms] \\
\midrule
DR-MPC & $-0.065$ $[-0.075,-0.055]$ & 0.45 & $+17.4$ & 1.41 \\
IF-SMPC & $+0.013$ $[+0.003,+0.022]$ & 14 & 0 & 1.51 \\
COV & $+0.001$ $[-0.008,+0.011]$ & 9.2 & $+0.7$ & 1.55 \\
W2 & $-0.042$ $[-0.052,-0.032]$ & 1.2 & $+14.6$ & 1.39 \\
ORC & $-0.012$ $[-0.022,-0.002]$ & 3.2 & $+1.1$ & 1.59 \\
\bottomrule
\end{tabular}
\end{table}

\section{Conclusion}\label{sec:conclusion}

We extended indirect-feedback stochastic MPC to a distributionally robust approach with a Gelbrich ambiguity set on the disturbance mean and covariance. The controller is recursively feasible and satisfies each CVaR constraint in closed loop for every disturbance law in the set, and the online problem remains a quadratic program. In the quadrotor study, it kept the empirical CVaR below the limit in all tested cases, while the moment-based baseline exceeded it in some, at a cost about 17--18\% higher. Future work includes disturbances that are correlated over time and a data-driven choice of the radius.\looseness=-1

\bibliographystyle{IEEEtran}
\bibliography{refs}

\end{document}